\documentclass[showpacs,aps,graphicx,twocolumn]{revtex4}
\usepackage{graphicx}

\begin{document}

\title{ Deterministic photonic spatial-polarization hyper-controlled-not gate assisted by  quantum dot inside one-side optical microcavity
\footnote{Published in Laser Phys. Lett. \textbf{10} (2013) 095202}}

\author{Bao-Cang Ren, Hai-Rui Wei and Fu-Guo Deng\footnote{Corresponding author: fgdeng@bnu.edu.cn} }

\address{Department of Physics, Applied Optics Beijing Area Major Laboratory,
Beijing Normal University, Beijing 100875, China}

\date{\today }

\begin{abstract}

Up to now, all the works about constructing quantum logic gates, an
essential part in quantum computing, are focused on operating on one
degree of freedom (DOF) of quantum systems. Here, we investigate the
possibility to achieve a scalable photonic quantum computing based
on two DOFs of quantum systems and construct a deterministic
hyper-controlled-not (hyper-CNOT) gate operating on both
spatial-mode and polarization DOFs of a photon pair simultaneously,
by using the giant optical Faraday rotation induced by a
single-electron spin in a quantum dot inside a one-side optical
microcavity as a result of cavity quantum electrodynamics. With this
hyper-CNOT gate and linear optical elements, two-photon four-qubit
cluster entangled states can be prepared and analyzed, which gives
an application to manipulate more information with less resources.
We analyze the experimental feasibility of this hyper-CNOT gate and
show that it
can be implemented with current technology.

\end{abstract}
 \maketitle

\section*{\uppercase\expandafter{\romannumeral1.}
Introduction}

Quantum mechanics theory, the core of quantum information, can
largely improve the power of dealing and transmitting information.
Quantum computing \cite{QC} is required for the precise control and
manipulation of the states of quantum systems in quantum information
process. It is proven that two-qubit controlled-not (CNOT) gates (or
the equivalent two-qubit quantum gates) assisted by single-qubit
gates are sufficient for universal quantum computing, so it is
important to construct a two-qubit CNOT gate. By far, many works
have been done on constructing a two-qubit  CNOT gate or its
equivalent in notably trapped ions \cite{NTI}, nuclear magnetic
spins \cite{NMS}, free electrons \cite{FE}, and polarized photons
\cite{KLM,linear2,linear3,linear5,linear6,nonlinear}. The pioneering
work by Knill, Laflamme, and Milburn  \cite{KLM} showed that  a
photonic CNOT gate could be created with the maximal success
probability of 3/4, resorting to only single-photon sources,
detectors, and linear optical elements such as beam splitters. Since
this original work, there has been a significant progress
\cite{linear2,linear3,linear5,linear6} in constructing a CNOT gate
with linear optics, which realizes probabilistically the nonlinear
coupling between two photons by using interference, at least two
ancilla photons, single-photon detectors, and conditioning. With
nonlinear optics, a deterministic CNOT gate can also be constructed.
In 2004, Nemoto and Munro \cite{nonlinear} constructed a near
deterministic CNOT gate using several single photon sources,
linear-optical elements, photon number resolving quantum
nondemolition detectors, and feed-forward. The key element of the
CNOT gate in their proposal is the quantum nondemolition detector
(QND) based on cross-Kerr nonlinearity.

Although there are many valuable works on constructing quantum
logical gates, especially CNOT gates, they  are all focused on
operating on one degree of freedom (DOF) of quantum systems. There
are, by far,  no quantum entangling gates operating on more than one
DOF of quantum systems. In fact, there are many advantages for
dealing with quantum information process in a larger Hilbert space,
especially for its robustness against noise \cite{noi} and the high
channel capacity. High-dimensional entanglement has been realized in
multipartite \cite{mulp1,mulp4} and multidimensional
\cite{muld1,muld4} quantum systems. Although a universal quantum
computing can be realized with two-qubit CNOT gates and single-qubit
operations, it will be convenient to have multi-qubit quantum logic
gates in quantum computation. In 2006, Fiur\'{a}\v{s}ek \cite{Fiu1}
proposed some schemes for the probabilistic direct realization of
the fundamental Toffoli and Fredkin gates with linear optics, and he
presented a scheme for linear optical quantum Fredkin gate based on
the combination of experimentally demonstrated linear optical
partial-SWAP gate and controlled-Z gates in 2008 \cite{Fiu2}. Gong
\emph{et al} \cite{Gon} also discussed the realization of a quantum
Fredkin gate by using CNOT gates with only linear optics and single
photons. The large Hilbert space with more than one DOF has also
been discussed in some applications in quantum communication in the
past few years \cite{mul1,mul3,mul5,mul6,mul7,mul8}, especially for
hyperentanglement which is defined as quantum systems entangled in
more than one DOF. Besides the task in which hyperentanglement is
used to assist quantum information processing in one  DOF, the
complete analyzer for hyperentangled Bell states has also been
constructed \cite{HBSA} with cross-kerr nonlinearity to increase the
channel capacity of long-distance quantum communication in more than
one DOF.

In this Letter, we  investigate the possibility of achieving
scalable photonic quantum computing based on two DOFs of quantum
systems, which is different from all the existing works about
constructing quantum logic gates operating on one DOF of quantum
systems. We construct a deterministic hyper-controlled-not
(hyper-CNOT) gate which is used to perform a CNOT gate on both the
spatial-mode and the polarization DOFs of a two-photon system
simultaneously, assisted by two quantum dots embodied in one-side
optical microcavities (QD-cavity). Exploiting the giant optical
Faraday rotation of the left-circularly and the right-circularly
polarized photons induced by a single-electron spin in a QD-cavity
system, we construct a four-qubit controlled-Z (CZ) gate by using
the two electron spins as the control qubits and the polarization
and the spatial mode of a photon as the two target qubits,
respectively. After the second photon interacts with the two
QD-cavity systems, the two electron spins are detected, and a
deterministic hyper-CNOT gate could be constructed with feed-forward
single-qubit operations. As an application of this hyper-CNOT gate,
one can use it to prepare entangled two-photon four-qubit cluster
states easily and analyze the 16 hyperentangled cluster states
simply, resorting to some linear optical elements. We analyze the
experimental feasibility of this hyper-CNOT gate, and our result
shows that it can be implemented with current technology.

\section*{\uppercase\expandafter{\romannumeral2.} Construction of deterministic spatial-polarization hyper-controlled-not gate}




In 2008, Hu  \emph{et al }\cite{QD1,QD2}   pointed out that the
interaction of left-circularly and right-circularly polarized
photons with a QD-cavity system can be used in quantum information
process. With this optical property of QD-cavity systems,
multi-qubit entangler \cite{QD1,QD2,QD3} and photonic polarization
Bell-state analyzer \cite{QD4,QD5} can be constructed. In 2010,
Bonato \emph{et al} \cite{QD5} constructed a CNOT gate operating on
hybrid quantum system composed of a photon and an electron spin
using the interface between the photon and the electron spin in  a
double-sided QD-cavity system in the weak coupling regime.

\begin{figure}[htbp]             
\centering\includegraphics[width=7.8 cm]{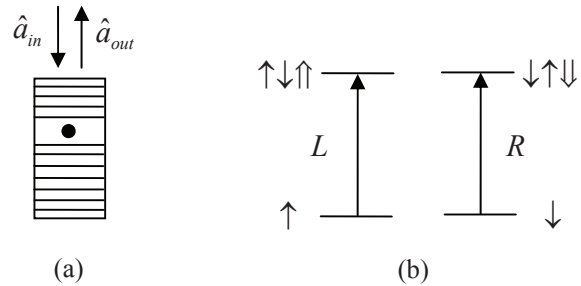} \caption{ The
$X^-$ spin-dependent transitions with circularly polarized photons.
(a) A charged QD inside a one-side micropillar microcavity
interacting with circularly polarized photons. (b) $X^-$
spin-dependent optical transition rules due to the Pauli's exclusion
principle. $L$ and $R$ represent a left-circularly and a
right-circularly polarized photons, respectively. $\uparrow$ and
$\downarrow$ represent the  spins of the excess electron.
$\uparrow\downarrow\Uparrow$ and $\downarrow\uparrow\Downarrow$
represent negatively charged exciton $X^-$.} \label{figure1}
\end{figure}

The QD-cavity system used in our proposal is constructed by a singly
charged QD (e.g., a self-assembled In(Ga)As QD or a GaAs interface
QD) located in the center of a one-side optical resonant cavity (the
bottom distributed Bragg reflectors is 100\% reflective and the top
distributed Bragg reflectors is partially reflective) to achieve the
maximal light-matter coupling. According to Pauli's exclusion
principle, a negatively charged exciton ($X^-$) consisted of two
electrons bound to one hole \cite{QD8} can be optically excited when
an excess electron is injected into the QD. The optical resonance of
$X^-$  with circularly  polarized  photons depends on the excess
electron spin in the QD \cite{QD9}, shown in Fig.\ref{figure1}. For
the excess electron-spin state $|\uparrow\rangle$, the negatively
charged exciton $|\uparrow\downarrow\Uparrow\rangle$ with two
electron spins antiparallel is created by resonantly absorbing a
left-circularly polarized photon $|L\rangle$. Here
$|\Uparrow\rangle$ describes the heavy-hole spin state
$|+\frac{3}{2}\rangle$.  For the excess electron spin
$|\downarrow\rangle$, the other negatively charged exciton
$|\downarrow\uparrow\Downarrow\rangle$ can be created by resonantly
absorbing a right-circularly polarized photon $|R\rangle$. Here
$|\Downarrow\rangle$ describes the heavy-hole spin state
$|-\frac{3}{2}\rangle$. This optical process can be described by
Heisenberg equations for the cavity field operator $\hat{a}$ and
$X^-$ dipole operator $\hat{\sigma}_-$ in the interaction picture
\cite{QD10},
\begin{eqnarray}                           \label{eq.1}
& & \frac{d\hat{a}}{dt}=-[i(\omega_c-\omega)+\frac{\kappa}{2}+\frac{\kappa_s}{2}]\,\hat{a}-g\,\hat{\sigma}_--\sqrt{\kappa}\,\hat{a}_{in}, \nonumber\\
& &\frac{d\hat{\sigma}_-}{dt}=-[i(\omega_{X^-}-\omega)+\frac{\gamma}{2}]\,\hat{\sigma}_- - g\,\hat{\sigma}_z\,\hat{a},\nonumber\\
& & \hat{a}_{out}=\hat{a}_{in}+\sqrt{\kappa}\,\hat{a},
\end{eqnarray}
where $g$ represents the coupling strength between the cavity mode
and $X^-$. $\kappa/2$ and $\kappa_s/2$ represent the decay rate and
the side leakage rate of the cavity field, respectively. $\gamma/2$
represents the decay rate of $X^-$. $\omega_{X^-}$, $\omega$, and
$\omega_c$ represent the frequencies of the  $X^-$ transition, the
input probe light, and the cavity mode, respectively.

If $X^-$ stays in the ground state at the most time (i.e.,
$\langle\sigma_z\rangle=-1$), the reflection coefficient of a
QD-cavity system for this weak excitation condition is \cite{QD1}
\begin{eqnarray}                           \label{eq.2}
r(\omega)=1-\frac{\kappa[i(\omega_{X^-}-\omega)+\frac{\gamma}{2}]}{[i(\omega_{X^-}-\omega)
+\frac{\gamma}{2}][i(\omega_c-\omega)+\frac{\kappa}{2}+\frac{\kappa_s}{2}]+g^2}.\nonumber\\
\end{eqnarray}
In the condition $g=0$, the QD is uncoupled to the cavity (the cold
cavity), and the reflection coefficient becomes
\begin{eqnarray}                           \label{eq.3}
r_0(\omega)=\frac{i(\omega_c-\omega)-\frac{\kappa}{2}+\frac{\kappa_s}{2}}{i(\omega_c-\omega)+\frac{\kappa}{2}+\frac{\kappa_s}{2}}.
\end{eqnarray}

Now, the optical process based on the $X^-$ spin-dependent
transitions is obtained. For a light in the state $|L\rangle$, a
phase shift of $\varphi_h$ is gotten for a hot cavity (the QD is
coupled to the cavity) with the excess electron spin in the state
$|\uparrow\rangle$, and a phase shift of $\varphi_0$ is gotten for a
cold cavity with the excess electron spin in the state
$|\downarrow\rangle$. For a light in the state $|R\rangle$, a phase
shift of $\varphi_0$ is gotten for a cold cavity with the excess
electron spin in the state $|\uparrow\rangle$, and a phase shift of
$\varphi_h$ is gotten for a hot cavity with the excess electron spin
in the state $|\downarrow\rangle$. By adjusting the frequencies of
the light ($\omega$) and the cavity mode ($\omega_c$), the
reflection coefficient can reach $|r_0(\omega)|\cong1$ and
$|r_h(\omega)|\cong1$ when the cavity side leakage is negligible. If
the linearly polarized probe beam in the state
$\frac{1}{\sqrt{2}}(|R\rangle+|L\rangle)$ is put into a one-side
QD-cavity system in the superposition spin state
$\frac{1}{\sqrt{2}}(|\uparrow\rangle+|\downarrow\rangle)$, the state
of system composed by the light and the electron spin after
reflection is
\begin{eqnarray}                           \label{eq.4}
\frac{1}{2}(|R\rangle+|L\rangle)\otimes(|\uparrow\rangle + |\downarrow\rangle)\rightarrow \frac{1}{2}e^{i\varphi_0}[(|R\rangle
 +e^{i\Delta\varphi}|L\rangle)\nonumber\\
|\uparrow\rangle+(e^{i\Delta\varphi}|R\rangle+|L\rangle)|\downarrow\rangle],
\end{eqnarray}
where $\Delta\varphi=\varphi_h-\varphi_0$,
$\varphi_0=arg[r_0(\omega)]$, and $\varphi_h=arg[r_h(\omega)]$.
$\theta_F^\uparrow=(\varphi_0-\varphi_h)/2=-\theta_F^\downarrow$ is
a Faraday rotation angle. After reflection, the light and the spin
become entangled with the different phase shifts of $|L\rangle$ and
$|R\rangle$ photons.

\begin{figure}[!h]
\centering\includegraphics[width=8.2 cm,angle=0]{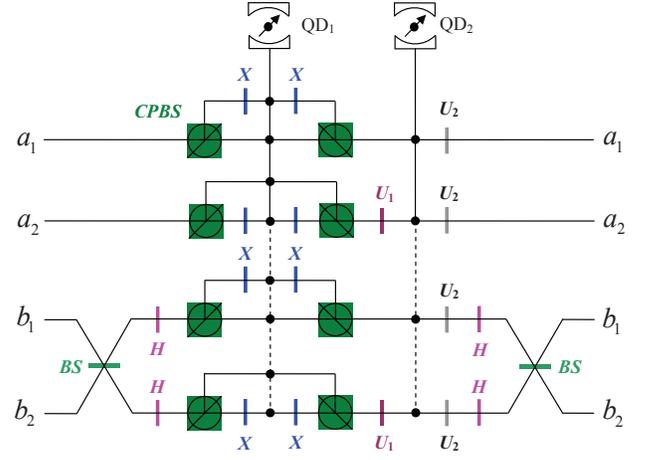}
\caption{(Color online)  Schematic diagram of our
spatial-polarization hyper-controlled-not gate operating on both the
spatial-mode and polarization degrees of freedom of a two-photon
system simultaneously. BS represents a 50:50 beam splitter which is
used to perform a Hadamard operation on the spatial-mode states of a
photon. CPBS represents a polarizing beam splitter in the circular
basis, which transmits the photon in the right-circular polarization
$\vert R\rangle$ and reflects the photon in the left-circular
polarization $\vert L\rangle$, respectively. $X_j$ represents a
half-wave plate (HWP$_1$) which is used to perform a polarization
bit-flip operation $X_j=|R\rangle\langle L|+|L\rangle\langle R|$
($j=1,2,...,8$). $H_k$ represents a half-wave plate (HWP$_2$) which
is used to perform a polarization Hadamard operation [$\vert
R\rangle \leftrightarrow \frac{1}{\sqrt{2}}(\vert R\rangle + \vert
L\rangle)$ and $\vert L\rangle \leftrightarrow
\frac{1}{\sqrt{2}}(\vert R\rangle - \vert L\rangle)$]($k=1,2,3,4$).
$U_{1m}$ represents a wave plate (WP$_1$) which is used to perform a
polarization phase-flip operation $U_{1m}=-i|R\rangle\langle
R|-i|L\rangle\langle L|$ to achieve the relative phase shift $-i$
between the two spatial modes of a photon ($m=2,4$). $U_{2n}$
represents a wave plate (WP$_2$) which is used to perform a
polarization phase-flip operation $U_{2n}=|R\rangle\langle
R|-i|L\rangle\langle L|$ ($n=1,2,3,4$). } \label{figure2}
\end{figure}

The principle of our deterministic spatial-polarization hyper-CNOT
gate for a two-photon system in two DOFs with the optical property
of one-side QD-cavity systems is shown in Fig.\ref{figure2}, which
can be used to perform a CNOT gate operation on both the spatial
mode and the polarization DOFs of a two-photon system by flipping
the spatial-mode target qubit and the polarization target qubit if
the two control qubits are in the spatia mode $|i_2\rangle$ and in
the polarization state $|L\rangle$, respectively. The QD$_1$ and
QD$_2$ are used to operate on the spatial-mode and the polarization
DOFs, respectively. By adjusting the frequencies
$\omega-\omega_c\approx\kappa/2$ to get the phase shifts of the left
and the right circularly polarized photons as $\varphi_0=-\pi/2$ and
$\varphi_h=0$, the interaction of a single photon with a QD-cavity
system can be described as
\begin{eqnarray}                           \label{eq.5}
&&|R,\uparrow\rangle\rightarrow-i|R,\uparrow\rangle,\;\;\;\;\;\;\; \;|L,\uparrow\rangle\rightarrow|L,\uparrow\rangle,\nonumber\\
&&|R,\downarrow\rangle\rightarrow|R,\downarrow\rangle,\;\;\;\;\;\;\;\;\;\;\;
\;|L,\downarrow\rangle\rightarrow-i|L,\downarrow\rangle.
\end{eqnarray}

\bigskip

(1) \emph{ Three-particle four-qubit hybrid controlled-Z gate}

\bigskip

We assume the initial states of the excess electron spin in QD$_1$
and the photon $a$ are
$\frac{1}{\sqrt{2}}(i|\uparrow\rangle+|\downarrow\rangle)_{e_1}$ and
$\vert \Phi_a\rangle_0 \equiv
(\alpha_1|R\rangle+\alpha_2|L\rangle)_a\otimes(\gamma_1|a_1\rangle+\gamma_2|a_2\rangle)$,
respectively. Here $|a_1\rangle$ and $|a_2\rangle$ represent the two
spatial modes of the photon $a$. After the photon $a$ passes through
the circular polarizing beam splitter CPBS (CPBS$_1$ and CPBS$_3$)
and the half-wave plate HWP$_1$   (X$_1$ and X$_3$), interacts with
the electron spin in QD$_1$, and then passes through another HWP$_1$
(X$_2$ and X$_4$), CPBS (CPBS$_2$ and CPBS$_4$) and the wave plate
WP$_1$ (U$_{12}$) in Fig.\ref{figure2}, the state of the system
composed of QD$_1$ and the photon $a$ is changed from $\vert
\Phi_{ae_1}\rangle_0 $ to $\vert \Phi_{ae_1}\rangle_1$. Here
\begin{eqnarray}                           \label{eq.6}
\vert \Phi_{ae_1}\rangle_0 &=& (\alpha_1|R\rangle+\alpha_2|L
\rangle)_a\otimes(\gamma_1|a_1\rangle+\gamma_2|a_2\rangle)
\nonumber\\ && \otimes \frac{1}{\sqrt{2}}(i|\uparrow\rangle
+|\downarrow\rangle)_{e_1},\nonumber\\
\vert \Phi_{ae_1}\rangle_1 &=&
\frac{1}{\sqrt{2}}(\alpha_1|R\rangle+\alpha_2|L\rangle)_a\otimes
[|\uparrow\rangle_{e_1}(\gamma_1|a_1\rangle\nonumber\\
&& \;\;\;\; +\,
\gamma_2|a_2\rangle)+|\downarrow\rangle_{e_1}(\gamma_1|a_1\rangle-\gamma_2|a_2\rangle)].
\end{eqnarray}
This is the result of the controlled-Z gate constructed by using the
electron spin $e_1$ as the control qubit and the spatial modes of
the  photon $a$ as the target qubit.

The initial state of QD$_2$ in Fig.\ref{figure2} is prepared to be
$\frac{1}{\sqrt{2}}(i|\uparrow\rangle+|\downarrow\rangle)_{e_2}$,
and the frequencies of the input photon and the cavity mode are
adjusted to be $\omega-\omega_c\approx\kappa/2$, as the same as
those for QD$_1$. After the photon $a$ interacts with QD$_2$ and
passes through WP$_2$ (U$_{21}$ and U$_{22}$) as shown in
Fig.\ref{figure2}, the state of the complicated system composed of
QD$_1$, QD$_2$, and the photon $a$ is changed from $\vert
\Phi_{ae_1e_2}\rangle_1$  to $\vert \Phi_{ae_1e_2}\rangle_2$. Here
\begin{eqnarray}                           \label{eq.7}
\vert \Phi_{ae_1e_2}\rangle_1&=&\vert \Phi_{ae_1}\rangle_1\otimes
\frac{1}{\sqrt{2}}(i|\uparrow\rangle
+|\downarrow\rangle)_{e_2},\nonumber\\
\vert
\Phi_{ae_1e_2}\rangle_2&=&\frac{1}{2}[|\uparrow\rangle_{e_2}(\alpha_1|R\rangle+\alpha_2|L\rangle)_a
+|\downarrow\rangle_{e_2}(\alpha_1|R\rangle\nonumber\\
&&-\alpha_2|L\rangle)_a]\otimes[|\uparrow\rangle_{e_1}(\gamma_1|a_1\rangle+\gamma_2|a_2\rangle)\nonumber\\
&&+|\downarrow\rangle_{e_1}(\gamma_1|a_1\rangle-\gamma_2|a_2\rangle)].
\end{eqnarray}
This is the result of the three-particle four-qubit hybrid
controlled-Z gate constructed by using the two electron spins as the
control qubits and the polarization and the spatial modes of the
photon $a$ as the two target qubits. That is, when the spin of the
electron $e_1$ is in the state $\vert \downarrow\rangle_{e_1}$, the
spatial mode $\vert a_2\rangle$ of the photon $a$ obtains a phase
shift $\pi$, and when the spin of the electron $e_2$ is in the state
$\vert \downarrow\rangle_{e_2}$, the polarization mode $\vert
L\rangle$ of the photon $a$ obtains a phase shift $\pi$.

\bigskip

(2) \emph{Spatial-polarization hyper-CNOT gate}

\bigskip

The second photon $b$ is prepared initially in the state
$\vert\Phi_b\rangle_0 =
(\beta_1|R\rangle+\beta_2|L\rangle)_b\otimes(\delta_1|b_1\rangle+\delta_2|b_2\rangle)$.
After Hadamard operations are performed on both the spatial mode and
the polarization DOFs of the photon $b$ (by making the photon $b$
pass through $BS_1$, $H_1$ and $H_2$), the state
$\vert\Phi_b\rangle_0$ is changed to be $\vert\Phi_b'\rangle_0\equiv
(\beta'_1|R\rangle+\beta'_2|L\rangle)_b\otimes(\delta'_1|b_1\rangle+\delta'_2|b_2\rangle)$.
Here $\beta'_1=\frac{1}{\sqrt{2}}(\beta_1+\beta_2)$,
$\beta'_2=\frac{1}{\sqrt{2}}(\beta_1-\beta_2)$,
$\delta'_1=\frac{1}{\sqrt{2}}(\delta_1+\delta_2)$ and
$\delta'_2=\frac{1}{\sqrt{2}}(\delta_1-\delta_2)$. Then we perform
unitary operations on the two electron spins in QD$_1$ and QD$_2$,
which transform the states $|\uparrow\rangle$ and
$|\downarrow\rangle$ to
$\frac{1}{\sqrt{2}}(i|\uparrow\rangle+|\downarrow\rangle)$ and
$\frac{1}{\sqrt{2}}(i|\uparrow\rangle-|\downarrow\rangle)$,
respectively. Subsequently, we let the photon $b$ pass through CPBS
(CPBS$_5$ and CPBS$_7$), HWP$_1$ (X$_5$ and X$_7$), QD$_1$, HWP$_1$
(X$_6$ and X$_8$), CPBS (CPBS$_6$ and CPBS$_8$), WP$_1$ (U$_{14}$),
QD$_2$ and WP$_2$ (U$_{23}$ and U$_{24}$). After the interaction
between the photon $b$ and the two QDs, the state of system composed
of QD$_1$, QD$_2$, and the photons $a$ and $b$ is changed from
$\vert \Psi_s\rangle_1$ to $\vert \Psi_s\rangle_2$. Here
\begin{eqnarray}                           \label{eq.8}
\vert \Psi_s\rangle_1 &= & \vert
\Phi_{ae_1e_2}\rangle_2\otimes(\beta'_1|R \rangle+\beta'_2|L \rangle)_b (\delta'_1|b_1\rangle+\delta'_2|b_2\rangle),\nonumber\\
\vert \Psi_s\rangle_2 &=&
[|\uparrow\rangle_{e_1}\gamma_1|a_1\rangle(\delta'_1|b_1\rangle+\delta'_2|b_2\rangle)
+ |\downarrow\rangle_{e_1}\gamma_2|a_2\rangle
\nonumber\\
&&  \times(\delta'_1|b_1\rangle-\delta'_2|b_2\rangle)]\otimes  [|\uparrow\rangle_{e_2}\alpha_1|R\rangle_a(\beta'_1|R \rangle\nonumber\\
&& +\beta'_2|L \rangle)_b +
|\downarrow\rangle_{e_2}\alpha_2|L\rangle_a(\beta'_1|R\rangle-\beta'_2|L\rangle)_b].
\end{eqnarray}
By performing Hadamard operations on the two electron spins in
QD$_1$ and QD$_2$ again after the photon $b$ passes through WP$_2$
(U$_{23}$ and U$_{24}$), the state of  the complicated system
becomes
\begin{eqnarray}                           \label{eq.9}
\vert \Psi_s\rangle_3
&=&\frac{1}{2}\{|\uparrow\rangle_{e_1}[\gamma_1|a_1\rangle(\delta'_1|b_1\rangle+\delta'_2|b_2\rangle)
+ \gamma_2|a_2\rangle(\delta'_1|b_1\rangle
\nonumber\\
&&\; -
\delta'_2|b_2\rangle)]+|\downarrow\rangle_{e_1}[\gamma_1|a_1\rangle(\delta'_1|b_1\rangle+\delta'_2|b_2\rangle)
\nonumber\\
&&\; - \gamma_2|a_2\rangle(\delta'_1|b_1\rangle-\delta'_2|b_2\rangle)]\}\nonumber\\
&&\otimes\{|\uparrow\rangle_{e_2}[\alpha_1|R\rangle_a(\beta'_1|R\rangle+\beta'_2|L \rangle)_b+\alpha_2|L\rangle_a\nonumber\\
&&\;(\beta'_1|R\rangle - \beta'_2|L \rangle)_b] + |\downarrow\rangle_{e_2}[\alpha_1|R\rangle_a(\beta'_1|R\rangle\nonumber\\
&&\;+\beta'_2|L \rangle)_b - \alpha_2|L\rangle_a(\beta'_1|R
\rangle-\beta'_2|L\rangle)_b]\}.
\end{eqnarray}
At last, we perform Hadamard operations on both the spatial-mode
(BS$_2$) and the polarization (H$_3$ and H$_4$) DOFs of the photon
$b$, the state $\vert \Psi_s\rangle_3$ is transformed into
\begin{eqnarray}                           \label{eq.10}
\vert \Psi_s\rangle_4
&=&\frac{1}{2}\{|\uparrow\rangle_{e_1}[\gamma_1|a_1\rangle(\delta_1|b_1\rangle+\delta_2|b_2\rangle)
+ \gamma_2|a_2\rangle(\delta_2|b_1\rangle
\nonumber\\
&&\; +
\delta_1|b_2\rangle)]+|\downarrow\rangle_{e_1}[\gamma_1|a_1\rangle(\delta_1|b_1\rangle+\delta_2|b_2\rangle)
\nonumber\\
&&\; - \gamma_2|a_2\rangle(\delta_2|b_1\rangle+\delta_1|b_2\rangle)]\}\nonumber\\
&&\otimes\{|\uparrow\rangle_{e_2}[\alpha_1|R\rangle_a(\beta_1|R\rangle+\beta_2|L \rangle)_b+\alpha_2|L\rangle_a\nonumber\\
&&\;(\beta_2|R\rangle + \beta_1|L \rangle)_b] + |\downarrow\rangle_{e_2}[\alpha_1|R\rangle_a(\beta_1|R\rangle\nonumber\\
&&\; +\beta_2|L \rangle)_b- \alpha_2|L\rangle_a(\beta_2|R
\rangle+\beta_1|L\rangle)_b]\}.
\end{eqnarray}
By measuring the two excess electron spins $e_1$ and $e_2$ in the
orthogonal basis $\{|\uparrow\rangle, \, |\downarrow\rangle\}$, a
deterministic spatial-polarization hyper-CNOT gate can be
constructed with feed-forward operations.

\bigskip

(3) \emph{Feed-forward operations}

\bigskip

If an auxiliary photon with the initial state $\vert
\varphi_i\rangle=\frac{1}{\sqrt{2}}(|R\rangle+|L\rangle)$ is put
into an optical microcavity, the state of the system composed of the
photon and the electron spin after reflection is changed as follows:
\begin{eqnarray}                           \label{eq.11}
\frac{1}{\sqrt{2}}(|R\rangle+|L\rangle)|\uparrow\rangle &\rightarrow& \frac{1}{\sqrt{2}}(|R\rangle+i|L\rangle)|\uparrow\rangle,\nonumber\\
\frac{1}{\sqrt{2}}(|R\rangle+|L\rangle)|\downarrow\rangle
&\rightarrow&
\frac{1}{\sqrt{2}}(|R\rangle-i|L\rangle)|\downarrow\rangle.
\end{eqnarray}
By detecting the auxiliary photon with the orthogonal linear
polarization basis $\{\frac{1}{\sqrt{2}}(|R\rangle + i|L\rangle),
\frac{1}{\sqrt{2}}(|R\rangle - i|L\rangle) \}$, the excess electron
spin in QD can be read out. If the auxiliary photon is projected on
$\frac{1}{\sqrt{2}}(|R\rangle+i|L\rangle)$, the excess electron spin
is  in the state $|\uparrow\rangle$. If the auxiliary photon is
projected on $\frac{1}{\sqrt{2}}(|R\rangle-i|L\rangle)$, the excess
electron spin is in the state $|\downarrow\rangle$. If we perform an
addition sign change $|a_2\rangle\rightarrow-|a_2\rangle$ on the
photon $a$ for $|\downarrow\rangle_{e_1}$ in QD$_1$ and an addition
sign change $|L\rangle_a\rightarrow-|L\rangle_a$ on the photon $a$
for $|\downarrow\rangle_{e_2}$ in QD$_2$, the state of the system
composed of the photons $a$ and $b$ becomes
\begin{eqnarray}                           \label{eq.12}
\vert \Psi_{ab}\rangle_c &=&
[\gamma_1|a_1\rangle(\delta_1|b_1\rangle+\delta_2|b_2\rangle)
+\gamma_2|a_2\rangle(\delta_1|b_2\rangle\nonumber\\
&&+\delta_2|b_1\rangle)]\otimes[\alpha_1|R\rangle_a(\beta_1|R\rangle+\beta_2|L\rangle)_b\nonumber\\
&& +\alpha_2|L\rangle_a(\beta_1|L\rangle+\beta_2|R\rangle)_b].
\end{eqnarray}
One can see that there is a bit flip on the spatial mode of the
photon $b$ (the target qubit)  when the spatial mode of the photon
$a$ (the control qubit) is $\vert a_2\rangle$. Moreover, there is a
bit flip on the polarization of the photon $b$ when the polarization
of the photon $a$ is $\vert L\rangle_a$. That is, the outcome shown
in Eq.(\ref{eq.12}) is the hyper-CNOT gate operating on the
two-photon system on both its polarization and its spatial-mode
DOFs. Moreover, this hyper-CNOT gate works in a deterministic way in
principle.

\section*{\uppercase\expandafter{\romannumeral3.} Preparation of two-photon four-qubit cluster state}

\begin{figure}[!h]
\centering\includegraphics[width=7.5 cm,angle=0]{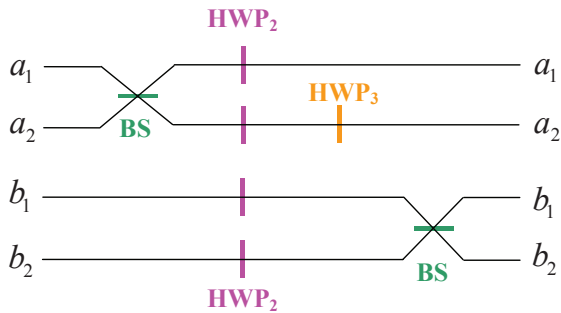}
\caption{(Color online)  Schematic diagram for the  preparation of a
two-photon four-qubit cluster state with a hyperentangled Bell
state. HWP$_2$ is used to perform a polarization Hadamard operation.
HWP$_3$ represents a half-wave plate which is used to perform a
polarization phase-flip operation $U_p=-|R\rangle\langle
R|+|L\rangle\langle L|$. BS represents beam splitter (50:50) which
is used to perform a spatial mode Hadamard operation.}
\label{figure3}
\end{figure}

As an application of our hyper-CNOT gate, we discuss the generation
and the complete analysis of two-photon four-qubit cluster entangled
states below. One can see that these tasks can be accomplished
easily and simply with a hyper-CNOT gate and  some linear-optical
elements.

With our hyper-CNOT gate operating on the spatial-mode and the
polarization DOFs of a photon pair in the initial state
$\frac{1}{2}[(|R\rangle+|L\rangle)_a(|a_1\rangle+|a_2\rangle)\otimes\vert
R\rangle_b|b_1\rangle]$, it is, in principle, easily to prepare the
hyperentangled two-photon four-qubit state
\begin{eqnarray}
\vert \Phi_{ab}\rangle_1=\frac{1}{2}(|R\rangle_a\vert
R\rangle_b+|L\rangle_a\vert L\rangle_b)\otimes(|a_1\rangle
|b_1\rangle + |a_2\rangle |b_2\rangle).\nonumber\\
\end{eqnarray}
With linear optical elements, this hyperentangled state can be
transformed into a two-photon four-qubit cluster state.

As shown in Fig.\ref{figure3}, after the  Hadamard operations
performed on both the polarization (HWP$_2$) and the spatial-mode
(BS) DOFs of the photon $a$, the hyperentangled Bell state becomes
\begin{eqnarray}                           \label{eq.13}
\vert\Phi_{ab}\rangle_2&=&  \frac{1}{4}[(|R\rangle + |L \rangle)_a |R\rangle_b + (|R\rangle - |L \rangle)_a|L\rangle_b]\otimes\nonumber\\
&&
[(|a_1\rangle+|a_2\rangle)|b_1\rangle+(|a_1\rangle-|a_2\rangle)|b_2\rangle].
\end{eqnarray}
Then we perform the spatial mode controlled polarization phase-flip
gate (HWP$_3$) on the photon $a$, and the two-photon state
$\vert\Phi_{ab}\rangle_2$ is transformed into
$\vert\Phi_{ab}\rangle_3$. Here
\begin{eqnarray}                           \label{eq.14}
\vert\Phi_{ab}\rangle_3
&=&\frac{1}{4}\{[(|R\rangle+|L \rangle) |a_1\rangle -(|R\rangle - |L \rangle) |a_2\rangle]_a|R\rangle_b|b_1\rangle\nonumber\\
&&+[(|R\rangle+|L \rangle) |a_1\rangle +(|R\rangle - |L \rangle) |a_2\rangle]_a|R\rangle_b|b_2\rangle\nonumber\\
&&+[(|R\rangle-|L \rangle) |a_1\rangle -(|R\rangle + |L \rangle) |a_2\rangle]_a|L\rangle_b|b_1\rangle\nonumber\\
&&+[(|R\rangle-|L \rangle) |a_1\rangle +(|R\rangle+|L
\rangle) |a_2\rangle]_a|L\rangle_b|b_2\rangle\}.\nonumber\\
\end{eqnarray}
After performing Hadamard operations on both the polarization
(HWP$_2$) and the spatial-mode (BS) DOFs of the photon $b$ at last,
we get the two-photon four-qubit cluster state
\begin{eqnarray}                           \label{eq.15}
\vert\Phi_{ab}\rangle_4
&=&\frac{1}{2}[|a_1\rangle|b_1\rangle(|R\rangle_a|R\rangle_b+|L\rangle_a|L\rangle_b)
\nonumber\\
&&-|a_2\rangle|b_2\rangle(|R\rangle_a|R\rangle_b-|L\rangle_a|L\rangle_b)].
\end{eqnarray}

From the process of the preparation of a two-photon four-qubit
cluster state, one can see that this state can be disentangled to
the hyperentangled Bell state $\vert \Phi_{ab}\rangle_1$ with some
Hadamard operations and a spatial-mode controlled polarization
phase-flip gate using linear optical elements. With a hyper-CNOT
gate, the 16 hyperentangled Bell states can be transformed into 16
two-photon four-qubit product states which can be distinguished
simply with linear optical elements and single-photon detectors.
That is, the 16 hyperentangled Bell states can be analyzed easily
with our spatial-polarization hyper-CNOT gate.

\section*{\uppercase\expandafter{\romannumeral4.} Discussion and Conclusion}

The spatial-polarization hyper-CNOT gate is constructed with the
interaction of circularly polarized photons and one-side QD-cavity
systems. According to Pauli's exclusion principle, this optical
property is caused by the different reflection phase shifts of the
left and the right circularly polarized photons. By adjusting the
frequencies as $\omega_c=\omega_{X^-}=\omega_0$,
$\omega-\omega_c\approx\kappa/2$ and the cavity side leakage rate as
$\kappa_s<1.3\kappa$ \cite{QD4}, the relative phase shift of
circularly polarized photons can achieve $\Delta\varphi=-\pi/2$.
Young \emph{et al} \cite{con} investigated the quantum-dot-induced
phase shift experimentally in 2011, and showed that a QD-induced
phase shift of 0.2 rad between an (effectively) empty cavity
($Q\sim51000,d=2.5\mu$m) and a cavity with a resonantly coupled QD
can be deduced using a single-photon level probe. The Hadamard
operation and unitary rotation operation of an electron spin can be
completed by single-spin rotations using nanosecond electron spin
resonance microwave pulses \cite{SSR}. The electron spin coherence
time can be extended to $\mu$s using spin echo techniques \cite{QD4}
to protect the  electron spin coherence with microwave pulses. The
optical coherence time of an exciton is ten times longer than the
cavity photon lifetime \cite{trion3}, with which the optical
dephasing only reduces the fidelity a few percent. The hole spin
dephasing is dominant in the spin dephasing of $X^-$, and it can be
safely neglected with the hole spin coherence time three orders
longer than the cavity photon lifetime \cite{trion6}. The
heavy-light hole mixing, which causes optical selection rule
unperfect, could be reduced by engineering the shape, size and type
of charged exciton \cite{QD4}.

\begin{figure}[htb]                    
\centering
\includegraphics[width=7.2 cm]{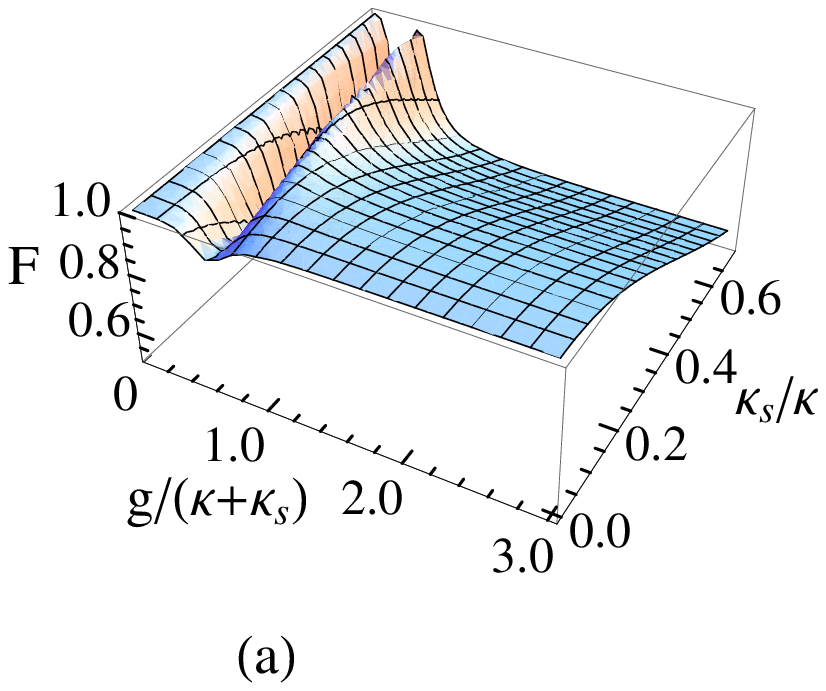}
\includegraphics[width=7.2 cm]{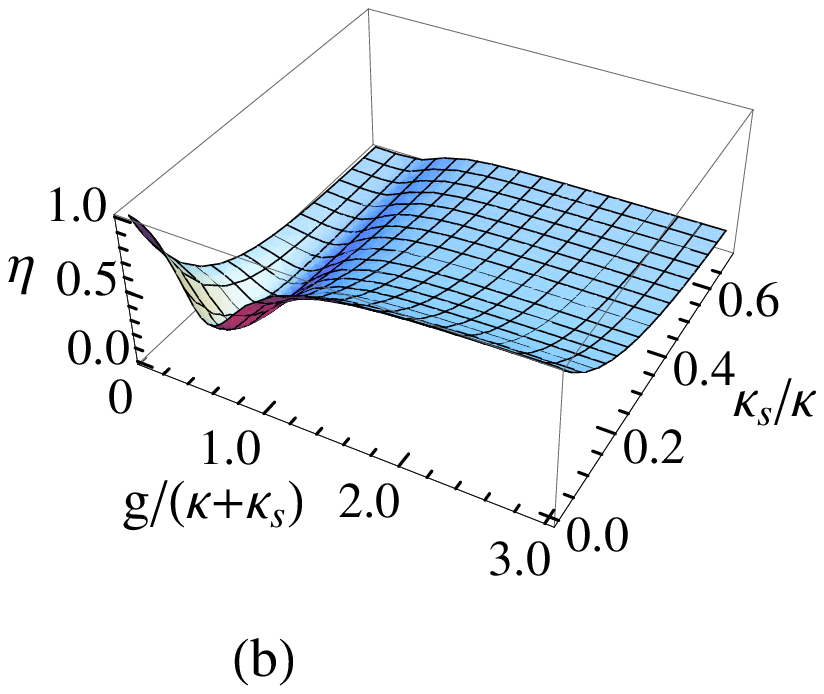}
\caption{(Color online)  Fidelity and efficiency of
spatial-polarization hyper-CNOT gate  vs the coupling strength and
side leakage rate with $\gamma=0.1\kappa$.} \label{figure4}
\end{figure}

In ideal conditions  in which one neglects the cavity side leakage,
the reflection coefficients are $|r_0(\omega)|\cong1$ and
$|r_h(\omega)|\cong1$, and the fidelity of our spatial-polarization
hyper-CNOT gate is nearly 100\%. Unfortunately, it is impossible to
neglect side leakage from the cavity, this being a rigorous
limitation to the QD¨Cmicropillar cavity in the experiment, and the
fidelity is reduced to $F=|\langle\psi_f|\psi\rangle|^2$, where
$|\psi\rangle$ and $|\psi_f\rangle$ describe the final states of an
ideal condition and a whole process considering external reservoirs,
respectively. The Faraday rotation angle is not very sensitive to
the side leakage when $\kappa_s<\kappa$ \cite{QD1}. Therefore, the
fidelity $F$ and the efficiency $\eta$ of  our spatial-polarization
hyper-CNOT gate are
\begin{eqnarray}                           \label{eq.16}
F&=&\frac{[2(|r_0|+|r_h|)^2]^4}{[2(||r_0|+|r_h||^2+||r_0|-|r_h||^2)(|r_0|^2+|r_h|^2)]^2}\nonumber\\
& &\times\frac{(|r_0|+|r_h|)^4}{(|r_0|^2+|r_h|^2)^2},\nonumber\\
\eta&=&[\frac{1}{2}(|r_0|^2+|r_h|^2)]^4.
\end{eqnarray}
The fidelity and the efficiency of our spatial-polarization
hyper-CNOT gate are shown in Fig. \ref{figure4}. One can see that
this hyper-CNOT gate  work efficiently in a strong coupling regime
($g>(\kappa+\kappa_s)/4$). It is challenging to achieve a strong
coupling in QD-cavity systems experimentally. The coupling strength
can be improved from $g\cong0.5(\kappa+\kappa_s)$ (the quality
factor $Q=8800$) \cite{couple} to $g\cong2.4(\kappa+\kappa_s)$
($Q\sim40000$) \cite{couple1} by engineering the sample designs,
growth  and fabrication in $d=1.5\mu$m micropillar microcavities. If
the coupling strength is $g\cong0.5(\kappa+\kappa_s)$, we can get
the fidelity and the efficiency as $F=94.3\%$ and $\eta=48.9\%$ for
$\kappa_s/\kappa=0$. If the coupling strength is
$g\cong2.4(\kappa+\kappa_s)$, we can get the fidelity and the
efficiency as $F=100\%$ and $\eta=96.3\%$ for $\kappa_s/\kappa=0$,
and $F=94.7\%$ and $\eta=47.3\%$ for $\kappa_s/\kappa=0.2$. Both the
fidelity and the efficiency become high with a strong coupling
strength, but they are reduced by the side leakage of cavity.

In experiment, the quality factor Q is dominated by the side leakage
and cavity loss rate in a micropillar, while $g$ is dominated by the
trion oscillator strength and the cavity modal volume. Hu \emph{et
al} \cite{QD4} pointed out that Q factor can be reduced by thinning
down the top mirror which can increase $\kappa$ and keep $\kappa_s$
in constant. Therefore, the coupling strength can be reduced to
$g\cong1.3(\kappa+\kappa_s)$ in a high-$Q$ micropillar
($Q\sim18900$) with the side leakage rate set to be
$\kappa_s/\kappa\sim0.2$, and the corresponding fidelity and the
efficiency are $F=96\%$ and $\eta=42.3\%$. It is quite demanded for
high efficiency operations to observe a small $\kappa_s/\kappa$ in a
strong coupling regime.

In conclusion, the construction of a deterministic
spatial-polarization hyper-CNOT gate can  be achieved  with the
giant optical Faraday rotation induced by a single-electron spin in
a quantum dot inside a one-side optical microcavity as a result of
cavity quantum electrodynamics. In order to obtain the giant optical
Faraday rotation, the frequencies of input photon  and cavity mode
should be adjusted to $\omega-\omega_c\approx\kappa/2$, and the side
leakage and the cavity loss rate $\kappa_s/\kappa$ should be
controlled as small as possible. With this  hyper-CNOT gate, the
preparation of two-photon four-qubit cluster states is easy, in
principle, and the complete analysis of hyperentangled Bell states
of two-photon systems is simple. We have analyzed the experimental
feasibility of this hyper-CNOT gate, concluding that it can be
implemented with current technology. This hyper-CNOT gate could give
the powerful capability for quantum computing and quantum
communication.

Certainly, we only discuss the construction of the
spatial-polarization hyper-CNOT gate by exploiting the nonlinear
optical property of one-side quantum dot-cavity systems. It can be,
in principle, constructed with other systems based on nonlinearity,
such as cross-Kerr media, nitrogen-vacancy centers, wave-guide
nanocavity, and so on. It is possible to construct multi-photon
quantum logical gates in the same way, which means a scalable
quantum computing based on two DOFs of photon systems is feasible.
Moreover, the present hyper-CNOT gate can be used to create
multi-photon hyperentangled states in more than one DOF and complete
the analysis on these multi-photon hyperentangled states as well.

\section*{ACKNOWLEDGEMENTS}

This work is supported by the National Natural Science Foundation of
China under Grant No. 11174039 and  NECT-11-0031.


\end{document}